\documentclass{article}
\pdfoutput=1
\usepackage{amsmath}
\usepackage{amssymb}
\usepackage{wasysym}
\usepackage{graphicx}
\usepackage{subfigure}
\usepackage{jheppub}

\title{Comment on ``Quantizing strings in de Sitter space''}

\author[a]{G. Alencar,}
\author[b]{M. O. Tahim,}
\author[a]{R. N. Costa Filho}
\affiliation[a]{Departamento de F\'{\i}sica, Universidade Federal do Cear\'{a},
Caixa Postal 6030, Campus do Pici, 60455-760, Fortaleza, Cear\'{a}, Brazil.}
\affiliation[b]{Universidade Estadual do Cear\'a, Faculdade de Educa\c{c}\~{a}o, Ci\^{e}ncias e Letras do Sert\~ao Central-
R. Epit\'acio Pessoa, 2554, 63.900-000 Quixad\'a, Cear\'a, Brazil.}
\emailAdd{geova@fisica.ufc.br}
\emailAdd{makarius.tahim@uece.br}
\emailAdd{rai@fisica.ufc.br}
\abstract{Trying to quantize the string in a de Sitter spacetime, Miao Li, Wei Song and Yushu Song presents a solution for the string equation of motion  (JHEP04(2007)042) that is not correct. This reopens the problem by them addressed.}


\begin{document}
\maketitle

\section{A Short Review of the Problem}

In the work \cite{Li:2007gf} string quantization in de Sitter spacetime is claimed to be performed. The main subject of that paper is to address the very early initial conditions of the universe in a cosmological context. The string viewpoint, together with field theory aspects of quantization on time dependent backgrounds, are the methods on which conclusions are based. Despite the importance of these themes, we would like to point that there is a mistake in the solution to the equation of motion for the problem presented. The starting point is the action for the bosonic string in a curved background that is given by
\begin{equation}\label{action}
S=-\frac{1}{4\pi\alpha'}\int d\tau\,d\sigma\,\sqrt{-h}\,h^{ab}\,\partial_{a}X^{\mu}\,\partial_{b}X^{\nu}\,G_{\mu\nu}(X)
\end{equation}
where $G_{\mu\nu}$ is the spacetime metric, $h_{ab}$ the worldsheet metric with $a,b$ running over $(\sigma,\tau)$ and $h=\det h_{ab}$. The background considered in Ref. \cite{Li:2007gf} is given by
 \begin{equation}\label{metric}
ds^{2}=-dt^{2}\,+\,e^{2Ht}\,(dx^{i})^{2}\ ;\,i=1,2,3.
\end{equation}
The gauge fixing is given by 
$\tau=t$, $h^{\tau\sigma}=0$, $-h=1$ and $h_{\sigma\sigma}(\sigma,t)$ is assumed to depend only on time ,{\it i.e.}, $h_{\sigma\sigma}(\sigma,t)=\omega(t)$. 

The equations of motion for the string are given by
\begin{eqnarray}\label{EOM1}
\partial_{t}(e^{2Ht}h_{\sigma\sigma}(t,\sigma)\partial_{t}X^{i}(t,\sigma))-e^{2Ht}\partial_{\sigma}(h_{\sigma\sigma}^{-1}(t,\sigma)\partial_{\sigma}X^{i}(t,\sigma))&=&0, \\
\partial_{t}h_{\sigma\sigma}(t,\sigma)+H\,e^{2Ht}\,[h_{\sigma\sigma}(t,\sigma)(\partial_{t}X^{i}(t,\sigma))^{2}-h_{\sigma\sigma}^{-1}(t,\sigma)(\partial_{\sigma}X^{i}(t,\sigma))^{2}]&=&0. \label{EOM2}
\end{eqnarray}
As can be seem above,  Eq. (\ref{EOM2}) is non-dynamical. However it is equivalent to the three constraints obtained by the variation of the action with respect to $h_{ab}$. With this the only equation left is (\ref{EOM2} ) which can be written as
\begin{equation}\label{EOM}
\partial_{t}(\eta^{-2}\partial_{t}X^{i}(t,\sigma))-\omega^{-2}\eta^{-2}\partial_{\sigma}^{2}X^{i}(t,\sigma))=0
\end{equation}
where $\eta=e^{-Ht}/\sqrt{\omega}$. 

The authors claim is that the general solution to equation (\ref{EOM}) is given by
\begin{eqnarray}
X^{i}(t,\sigma)&=&x_{0}+\int^{t}du\,\eta^{2}(u)\,p^{i}+\sum_{m\in Z/\{0\}}\eta(t)\Big[\frac{a_{m}^{i}(t)}{\sqrt{2|\lambda_{m}(t)|}}e^{-i\int^{t}du\lambda_{m}(u)}e^{im\sigma} \nonumber
\\
&+&\frac{\tilde{a}_{m}^{i}(t)}{\sqrt{2|\lambda_{m}(t)|}}e^{-i\int^{t}du\lambda_{m}(u)}e^{-im\sigma}\Big]\label{solution}
\end{eqnarray}
and
\begin{equation}\label{adot}
\dot{a}_{m}^{i}(t)=\frac{\dot{\lambda}_{m}(t)}{2\lambda_{m}(t)}\tilde{a}_{-m}^{i}(t)\,e^{2i\int^{t}du\lambda_{m}(u)},\;\dot{\tilde{a}}_{m}^{i}(t)=\frac{\dot{\lambda}_{m}(t)}{2\lambda_{m}(t)}\,a_{-m}^{i}(t)\,\,e^{2i\int^{t}du\lambda_{m}(u)},
\end{equation}
where
\begin{equation}\label{labda}
\lambda_{m}(t)=sgn(m)\sqrt{\frac{m^{2}}{\omega^{2}}-\eta\partial_{t}^{2}(\eta^{-1})}. 
\end{equation}
In the last equation $sgn(m)=1$ for $m>0$ and $sgn(m)=-1$ for $m<0$. After this all the results obtained in Ref. \cite{Li:2007gf} are based on Eq. (\ref{solution}): quantization, resolution of the constraints and finally the mass spectrum. In the next section we will comment about the above solution. 

\section{Checking the Solution}

In this section we show, by direct substitution, that (\ref{solution}) is not a solution of the string equations of motion (\ref{EOM}). For this we first define $X^{i}=\eta Y^{i}$ to obtain an equation for $Y^{i}$ given by
\begin{equation}\label{EOMY}
\partial_{t}^{2}Y^{i}=-[\eta\partial_{t}^{2}(\eta^{-1})+\omega^{-2}\eta^{-1}\partial_{\sigma}^{2}]Y^{i}.
\end{equation}
The first two terms of equation (\ref{solution}) satisfies equation (\ref{EOM}). Therefore we will focus on the modes with $m\neq0$. From now on $Y^{i}$ will refer to these modes and it is given by
\begin{equation}\label{Y}
Y^{i}(t,\sigma)=\sum_{m\in Z/\{0\}}[\frac{a_{m}^{i}(t)}{\sqrt{2|\lambda_{m}(t)|}}e^{-i\int^{t}du\lambda_{m}(u)}e^{im\sigma}+\frac{\tilde{a}_{m}^{i}(t)}{\sqrt{2|\lambda_{m}(t)|}}e^{-i\int^{t}du\lambda_{m}(u)}e^{-im\sigma}].
\end{equation}

Now we will substitute this in the right and left hand sides of the equation of motion (\ref{EOMY}) and show that they are different. The substitution on the RHS is more simple and gives us 
\begin{equation}\label{RHS}
-\sum_{m\in Z/\{0\}}\lambda_{m}^{2}[\frac{a_{m}^{i}(t)}{\sqrt{2|\lambda_{m}(t)|}}e^{-i\int^{t}du\lambda_{m}(u)}e^{im\sigma}+\frac{\tilde{a}_{m}^{i}(t)}{\sqrt{2|\lambda_{m}(t)|}}e^{-i\int^{t}du\lambda_{m}(u)}e^{-im\sigma}].
\end{equation}
For the LHS we first compute the first derivative of (\ref{Y}) to obtain
\begin{eqnarray}
&&\partial_{\tau}Y^{i}=\sum_{m\in\mathbb{Z}/\{0\}}\Big[ \frac{\tilde{a}_{-m}^{i}(t)\dot{|\lambda_{m}}|}{(2|\lambda_{m}(t)|)^{3/2}}e^{2i\int^{t}du\lambda_{m}(u)}-\frac{a_{m}^{i}(t)}{(2|\lambda_{m}(t)|)^{3/2}}\dot{|\lambda_{m}}|-i\frac{a_{m}^{i}(t)}{\sqrt{2|\lambda_{m}(t)|}}\lambda_{m}\Big] e^{-i\int^{t}du\lambda_{m}(u)}e^{im\sigma}\nonumber
\\
&+&\sum_{m\in\mathbb{Z}/\{0\}}\Big[ \frac{a_{-m}^{i}(t)\dot{|\lambda_{m}}|}{(2|\lambda_{m}(t)|)^{3/2}}e^{2i\int^{t}du\lambda_{m}(u)}-\frac{\tilde{a}_{m}^{i}(t)}{(2|\lambda_{m}(t)|)^{3/2}}\dot{|\lambda_{m}}|-i\frac{\tilde{a}_{m}^{i}(t)}{\sqrt{2|\lambda_{m}(t)|}}\lambda_{m}\Big] e^{-i\int^{t}du\lambda_{m}(u)}e^{-im\sigma}
\end{eqnarray}
where we have used (\ref{adot}). In the first and fourth terms above we perform the change $m\to-m$ and these terms will respectively cancel the fifth and second ones. We then get
\begin{equation}
\partial_{\tau}Y^{i}=-\sum_{m\in Z/\{0\}}i\lambda_{m}[\frac{a_{m}^{i}(t)}{\sqrt{2|\lambda_{m}(t)|}}e^{-i\int^{t}du\lambda_{m}(u)}e^{im\sigma}+\frac{\tilde{a}_{m}^{i}(t)}{\sqrt{2|\lambda_{m}(t)|}}e^{-i\int^{t}du\lambda_{m}(u)}e^{-im\sigma}]\label{dY}
\end{equation}
The second derivative can be obtained by a shortcut. Equation (\ref{dY}) above is just the multiplication of the modes of $Y^{i}$ by $-i\lambda_{m}(t)$. Therefore deriving the term in brackets will just gives us another factor of $-i\lambda_{m}$ and we get
\begin{eqnarray}
\partial_{\tau}^{2}Y^{i}=&&-\sum_{m\in Z/\{0\}}\lambda_{m}^{2}\Big[\frac{a_{m}^{i}(t)}{\sqrt{2|\lambda_{m}(t)|}}e^{-i\int^{t}du\lambda_{m}(u)}e^{im\sigma}+\frac{\tilde{a}_{m}^{i}(t)}{\sqrt{2|\lambda_{m}(t)|}}e^{-i\int^{t}du\lambda_{m}(u)}e^{-im\sigma}\Big]\nonumber \\
&-&\sum_{m\in Z/\{0\}}i\dot{\lambda}_{m}\Big[\frac{a_{m}^{i}(t)}{\sqrt{2|\lambda_{m}(t)|}}e^{-i\int^{t}du\lambda_{m}(u)}e^{im\sigma}+\frac{\tilde{a}_{m}^{i}(t)}{\sqrt{2|\lambda_{m}(t)|}}e^{-i\int^{t}du\lambda_{m}(u)}e^{-im\sigma}\Big].\label{LHS}
\end{eqnarray}
The first term of Eq. (\ref{LHS}) is equal to that of Eq. (\ref{RHS}) but the second can not be canceled. Therefore Eq. (\ref{solution}) is not a solution of the string equation of motion (\ref{EOM}). 

\section{Conclusion} 

It has been shown here that the presented solution, by simple inspection, do not satisfy the string equations of motion. We point that due to this, the conclusions made in Ref.  \cite{Li:2007gf} can not be the correct ones. One of main results is the mass spectrum given by
\begin{equation}
\alpha'M^{2}=4N+2E_{0}-\sum_{m,i}H^{2}\,(1+2N_{m}^{i}\,\tilde{N}_{m}^{i}+N_{m}^{i}+\tilde{N}_{m}^{i}).
\end{equation}
This differs from previous results found in Ref. \cite{deVega:1987veo} and the authors states that this may be due to the different gauge fixing. However this seems not to be the case. 

\section*{Acknowledgments}

The authors would like to thanks Alexandra Elbakyan and sci-hub, for removing all barriers in the way of science.
We acknowledge the financial support provided by Funda\c c\~ao Cearense de 
Apoio ao Desenvolvimento Cient\'\i fico e Tecnol\'ogico (FUNCAP)  through PRONEM PNE-0112-00085.01.00/16 and the Conselho 
Nacional de Desenvolvimento Cient\'\i fico e Tecnol\'ogico (CNPq).

\end{document}